\documentclass[11pt]{article}
\usepackage{graphicx}
\usepackage{cs12,html,hyperref}

\begin{document}
 
\title{Low-Mass Stars and Brown Dwarfs Around $\sigma$ Orionis}
 
\author{M. J. Kenyon\altaffilmark{1}, R. D. Jeffries\altaffilmark{1}, 
T. Naylor\altaffilmark{2}}
\altaffiltext{1}{Department of Physics, Keele University, UK}
\altaffiltext{2}{School of Physics, University of Exeter, UK}

\index{*$\sigma$ Orionis}
\index{*$\sigma$ Ori}
\index{Spectroscopy}

\begin{abstract}
We present optical spectroscopy of 71 photometric candidate low-mass
members of the cluster associated with ${\sigma}$ Orionis. Thirty-five 
of these are found to pass the lithium test and hence are confirmed as
true cluster members, covering a mass range of ${\leq0.055}$-$0.3\,M_{\sun}$, 
assuming a mean cluster age of ${\leq5}$\,Myr. We find 
evidence for an age spread on the ($I$, $I\!-\!J$) colour
magnitude diagram, members appearing to lie in the range 1-7\,Myr. There are, 
however, a significant fraction of candidates that are non-members, including 
some previously identified as members based on photometry alone. We see some 
evidence that the ratio of spectroscopically confirmed members to photometric 
candidates decreases with brightness and mass. This highlights the importance 
of spectroscopy in determining the true initial mass-function.
\end{abstract}

\section{Introduction}
$\sigma$ Orionis is the brightest member of a young ($<7$\,Myr) stellar cluster
of the same name, located in the Orion OB1b association at a distance modulus 
of $7.8$-$8.3$ (B\'ejar et al. 2001). Assuming that all members of this cluster
are coeval, it represents an ideal hunting ground for very low-mass stars and 
brown dwarfs, since such objects are at their most luminous when young. When 
combined with the fact that the cluster is unlikely to have undergone any 
dynamical evolution, such as evaporation or mass segregation, $\sigma$ Orionis
seems like an ideal candidate for studying the initial mass function.
The mass function can be determined from photometry but it is possible that 
contaminating objects, such as foreground M-dwarfs, bias the result. One way of 
eliminating these polluting objects from the colour magnitude diagram (CMD) is 
to perform a spectroscopic survey of cluster candidates. Since the cluster is 
very young all low-mass members should retain their initial lithium.

\section{Observations}
The INT wide-field camera was used to obtain $R$ and $I$ band photometry of a 
one square degree area of the $\sigma$ Orionis cluster during October 1999. 
Complete to $I\sim21$ and centered on the star after which the cluster is named,
the photometry was used to select $82$ cluster candidates for follow-up spectroscopy. Unfortunately, eleven of these targets were ``lost'' from the dataset 
because they fell down fibres of poor transmission (see below). The targets covered an $I$ 
magnitude range of $14.9$-$18.1$, the fainter end of this scale corresponding 
to an object with an age-dependent mass of $0.03$-$0.06M_{\sun}$ (at 
$1$-$7$\,Myr), i.e., just into the realm of brown dwarfs ($\la0.075M_{\sun}$). 
Figure $1$ shows the spectroscopic targets as triangles on an ($I$, \ri)
colour-magnitude diagram.

\begin{figure}[h]
\begin{center}
\includegraphics[width=7cm]{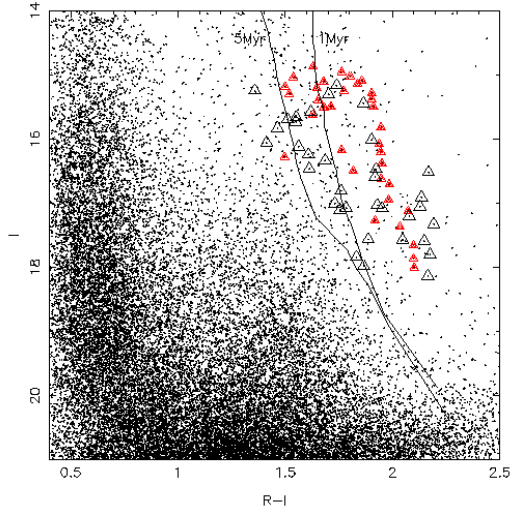}
\caption{(I, \ri) colour magnitude diagram for one square degree around $\sigma$
Orionis. Spectroscopic targets are shown as triangles - filled ones indicating 
cluster membership, open indicating no evidence of lithium. Baraffe et al. 
(1998) isochrones are also shown.}
\end{center}
\end{figure}

During December 1999 we obtained follow-up high-resolution ($1$\AA~FWHM) optical
spectroscopy for the photometric members using the Wide Field Fibre Optic 
Spectrograph (WYFFOS) on the $4.2$m William Herschel Telescope. The large-fibre
module was used ($2.7$\arcsec\,diameter fibres) with spectra being recorded via
the TEK 6 1024 pixel CCD. The $6625$\AA~echelle filter was selected, giving a 
spectral coverage of approximately $6400$-$6800$\AA, with a nominal dispersion 
of $0.4$\AA~pixel$^{-1}$. Spectra were taken over two nights and of the $71$ 
targets observed, $29$ were common to both of these runs. Total exposure times 
were; $4.25$ hours per target on the night of December $12$, and $4.33$ hours 
on December $13$. Arc spectra were obtained using a Tungsten lamp and sky 
exposures were taken following each target frame.

Reduction was carried out predominantly within the IRAF environment, 
using the WYFRED procedure. All images were initially bias subtracted by 
interpolating the overscan region across the whole of the frame. Flats were then
used to remove the instrumental signature of the CCDs. WYFRED employed an 
optimal method to extract the spectra from a frame, each spectrum 
was dispersion corrected and wavelength calibrated using a function 
determined from the arc spectra. Finally, a fibre-to-fibre throughput correction
was applied using offset-sky frames, and sky subtraction performed with mean 
sky spectra obtained with dedicated sky fibres. Spectra from the same night 
were co-added to increase the signal to noise ratio, but the decision was
made not to combine spectra from the two nights as this could mask any short 
period binaries.

\section{Lithium}
The ``lithium test'' exploits the fact that objects with masses 
$M\la0.065M_{\sun}$ never attain the high internal temperatures 
($\ga2.6\times10^6$K) required to burn lithium as a nuclear fuel. 
Stars above this mass-limit will consume the element, the age at which lithium 
destruction begins is determined by the mass of the star - the lower the mass, 
the longer until the onset of burning. The current work investigates the cluster
surrounding $\sigma$ Orionis, members of which are believed to have ages of 
$\la7$\,Myr. At this age, stars in the approximate mass range 
$0.4$-$0.7M_{\sun}$ will have consumed most of their initial lithium. All 
of our spectroscopic targets are believed to be less massive than 
$0.3M_{\sun}$, hence if the true cluster age is $\leq7$Myr, and there is 
little or no age spread, one would expect all the genuine cluster members to contain 
lithium.

Each spectrum was visually inspected for the characteristic $6707.8$\AA\, 
Lithium I doublet, and equivalent widths calculated. Briefly, thirty-five 
objects were found to contain significant lithium and are classified 
accordingly as definite cluster members. Assuming a cluster age of 
$\sim5$\,Myr, they range in mass from  $0.055$-$0.3M_{\sun}$.
A selection of these are shown in Figure 2, labelled with their $I$ magnitudes.
\begin{figure}[h]
\begin{center}
\includegraphics[width=7cm]{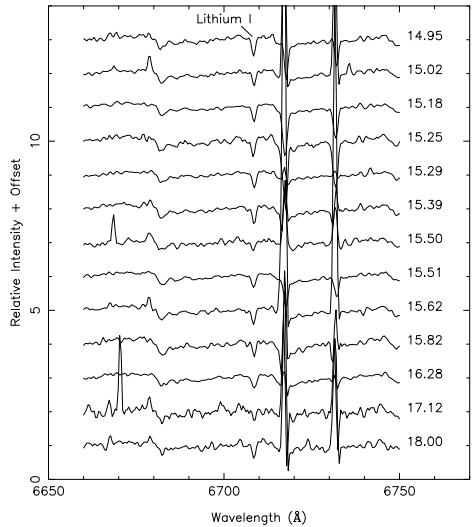}
\caption{Selection of cluster members' spectra, highlighting the Lithium I 
absorption feature at $6708$\AA. All spectra have been normalised and had a 
constant offset applied, labels are $I$ magnitudes}
\end{center}
\end{figure}

\section{Discussion}
Figure 1 shows an ($I$, \ri) CMD with the lithium members drawn as closed 
triangles and non-members as open triangles, isochrones are calculated from 
the Baraffe et al. (1998) models. There appears to be little correlation between
the isochrones and the observed cluster sequence. This seems to confirm the 
known mis-match between model calculations of \ri\, colours and their empirical 
counterparts. 2MASS $J$ band photometry was obtained for all but two of our 
targets, allowing us to construct an ($I$, $I-J$) CMD, see Figure 3.
Comparison between the identified cluster members and $I-J$ isochrones does 
provide a better match but suggests an age spread of $\sim1$-$7$\,Myr, even 
allowing for unresolved binary systems.

\begin{figure}[h]
\begin{center}
\includegraphics[width=7cm]{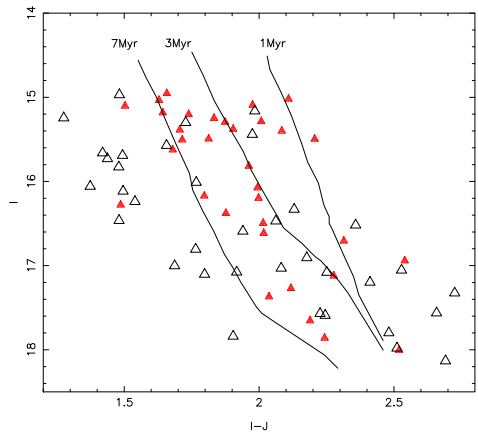}
\caption{($I,I\!-\!J$) colour magnitude diagram containing all but two of our $35$ 
cluster members (filled symbols), non members are indicated by open triangles. 
Isochrones are from Baraffe et al. (1998).}
\end{center}
\end{figure}

Figure 4 shows two histograms, the blue area representing the distribution of 
all $71$ spectroscopic targets in the survey, the red overlay corresponds to the
$35$ objects in which lithium was detected. The inference drawn from this figure
is that photometric selection alone is insufficient to identify cluster
members. In addition, the fraction of non-members appears to increase with 
magnitude. Thus a mass function based only on photometric criteria would 
over-estimate the numbers of brown dwarfs in the cluster with respect to higher
mass stars. It is important to note though, that of the $36$ objects classified as 
non-members, several have spectra which display a very small signal to 
noise ratio. This means that we cannot entirely rule out the presence of 
undepleted lithium within these objects, and consequently they may turn out to
be cluster members.

\begin{figure}[t]
\begin{center}
\includegraphics[width=7.5cm]{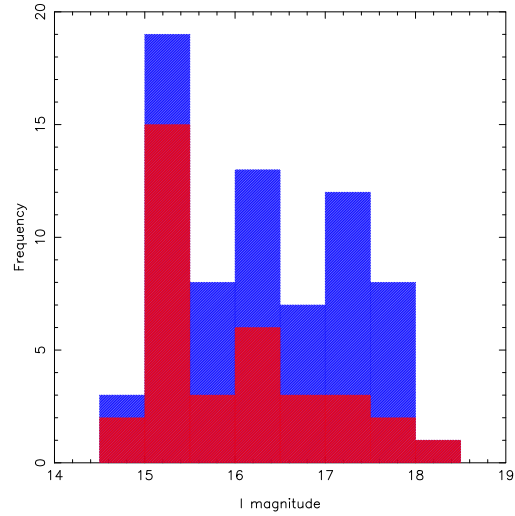}
\caption{Histogram of cluster membership as a function of $I$ magnitude. Red 
columns represent the number of cluster members found per magnitude bin, shown 
as a fraction of the total spectroscopic targets within that bin.}
\end{center}
\end{figure}

\begin{figure}[t]
\begin{center}
\includegraphics[width=7.5cm]{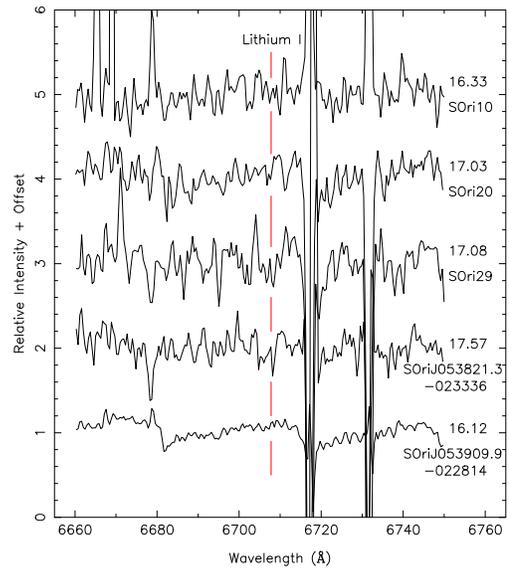}
\caption{Objects common to this survey and Bejar et al. (2001) which we find are
not cluster members, labelled with $I$ magnitudes and identifiers from Bejar et
al. (2001).}
\end{center}
\end{figure}
\clearpage

Bejar et al. (2001) present $64$ very low-mass $\sigma$ Orionis cluster member 
candidates using three colour ($I$, $Z$ and $J$) photometry. We are able to 
identify eighteen of their objects within our own survey, of these we are 
convinced that eight definitely show Lithium. However, five of their candidates 
show no clear evidence of cluster membership, these objects are all fainter than
$I=16$ and are shown in Figure 5. This goes some way to demonstrating the 
importance of backing up photometric cluster membership with spectroscopy.

We believe there is still much to be done with the dataset - current work 
involves determining radial velocities for all objects. This will provide a 
further discriminant between cluster members and contaminating objects, plus
allow the detection of binary systems. To date, we have found seven potential 
spectroscopic binary systems among the $35$ lithium-rich objects, with implied 
maximum orbital periods of a few days.

\end{document}